\documentclass[journal=cmatex,manuscript=article]{achemso}
\setkeys{acs}{articletitle=true}

\usepackage{charter,amssymb,mhchem,url}

\title{Metallic Oxides and the Overlooked Role of Bandwidth}

\author{Aurland\,K.\,Watkins} 
\affiliation{Materials Department and Materials Research Laboratory\\
University of California, Santa Barbara, California 93106, United States}

\author{Anthony\,K.\,Cheetham} 
\affiliation{Materials Department and Materials Research Laboratory\\
	University of California, Santa Barbara, California 93106, United States}

\author{Ram\,Seshadri} 
\affiliation{Materials Department and Materials Research Laboratory\\
	University of California, Santa Barbara, California 93106, United States}
\alsoaffiliation{Department of Chemistry and Biochemistry\\
	University of California, Santa Barbara, California 93106, United States}
\email{seshadri@mrl.ucsb.edu}

\begin{document}

\sloppy 

\clearpage

\begin{abstract}
Oxides exhibiting metallic conduction are crucial for various applications, including fuel cells, battery 
electrodes, resistive and magnetoresistive materials, electrocatalysts, transparent conductors, and 
high-temperature superconductors. Oxides that approach metallicity also play significant roles in 
switching applications, where the metal-insulator transition phenomenon is utilized across a range 
of technologies. This perspective, motivated by the question of when oxides are metallic, employs 
electronic structure calculations on metallic oxides to identify the typical feature in the electronic 
structures that promote metallic behavior. The critical factor of the bandwidth of the electronic energy 
bands near the Fermi energy is emphasized since it has been somewhat overlooked in the literature. For 
example, bandwidth considerations would suggest that the recently proposed phosphate ``LK-99'' would 
never be a suitable target for superconductivity. From the analysis performed here, we learn that if 
the width of the conduction band as obtained from density functional theory-based electronic structure 
calculations is less than 1\,eV, then the likelihood of obtaining a metallic compound is vanishingly 
small. This survey of representative oxide metals highlights the essential elements of extended 
covalency that lead to wide bands. A key takeaway is that oxyanion compounds such as borates, 
carbonates, silicates, sulfates, nitrates, and phosphates are unlikely to exhibit metallic conduction at 
ambient pressure. While the focus here is on oxides, the general findings should apply across various 
material families, extending to organic crystals, polymers, and framework materials.
\end{abstract}

\clearpage

\section{Introduction}

Titan, the largest moon of Saturn and the second largest moon in the solar system, has a dense atmosphere with 
\ce{N2}, \ce{CH4}, and \ce{H2} as the main components. We would anticipate that the mineral chemistry of Titan's crust 
is very different from the mineral chemistry seen on Earth.\cite{Titan2018} With an atmosphere that is relatively 
rich in \ce{O2}, and with oxophilic elements like Mg, Al, Si, Ti, and Fe dominating the Earth's crust, it is no 
surprise that oxides are ubiquitous, potentially suggesting an anthropic principle,\cite{Carter1974}  
for terrestrial mineralogy.  Oxides are often preferred over other material classes for a range of applications and 
operating conditions precisely because of the ambient environment in which they function. From a materials functionality 
perspective, oxides possess another virtue, poised as they are between being ionic and covalent in their extended 
interactions.\cite{Rao1989} This delicate balance allows some oxides to be highly insulating, with room-temperature 
resistivities as high as 10$^{20}$\,$\Omega$\,cm for fused \ce{SiO2}, while others are highly conducting, such 
as \ce{ReO3}, which displays resistivities lower than 10$^{-5}$\,$\Omega$\,cm at room temperature and 
is a better metal (displaying greater conductivity) than silver at low temperatures.\cite{Ferretti1965} 

Oxides displaying metallic conductivity or displaying an ability to be nudged into electronic conduction through 
doping or substitution, are important across a range of applications. A few examples include
the all-important Li–ion battery cathode LiCoO$_2$\cite{Mizushima1980}, which becomes metallic upon Li removal 
and concurrent Co oxidation\cite{Motohashi2009} enabling the high currents required in high performance 
applications. In contrast, the olivine-structured cathode material \ce{LiFePO4}\cite{Padhi1997} --- otherwise 
appealing for the relatively earth-abundant elemental components --- is  
insulating at all stages of lithiation and requires nanostructuring and coating with conductive carbon for effective
functioning.\cite{Goodenough2012} The effectiveness of Wadsley-Roth--structured oxide anode materials in 
conducting electronic charge when free d electrons are available greatly enhances their ability to perform at high 
rates.\cite{Griffith2018,Griffith2019,Preefer2020} One of the most effective electrocatalysts for the oxygen evolution
reaction is the metallic oxide \ce{IrO2}.\cite{McCrory2015} In solid-oxide fuel cells, the cathodes 
are often derived from cobalt-containing perovskites that combine some catalytic activity with efficient electronic and 
oxide ion conduction.\cite{Sun2010} The phenomenon of electroresistive switching, relating to so-called 
memristors,\cite{Strukov2008} is often seen in oxides that are proximal to metallic states.\cite{Watanabe2001}

Oxides are a playground for interesting low-temperature phenomena, starting historically with the discovery in 1939 
by Verwey of the now eponymous charge-ordering transition in magnetite \ce{Fe3O4}\cite{Verwey1939}, the
structural underpinnings of which took several decades to resolve.\cite{Senn2012} In 1950, Van Santen and
Jonker\cite{VanSanten1950} discovered concurrent magnetic ordering and metallic behavior in perovskite manganese 
oxides. Several decades later, the fascinating phenomena of colossal magnetoresistance was established in these
oxides.\cite{Ramirez1997} The metal--insulator transition seen upon cooling in oxides like \ce{VO2} and \ce{V2O3}
initiated a rich and expansive field of research on the topic. The role that oxides play in this field
can be gleaned from the authoritative review by Imada, Fujimori, and Tokura\cite{Imada1998} where much of the focus 
is on oxide material families. Arguably, some of the most interesting oxide metals are ones that transition to the 
superconducting state when cooled, and include spinel oxides of Ti\cite{Johnston1973}, perovskite oxides with 
Pb\cite{Sleight1975}, and Bi\cite{Mattheiss1988}, and the famous layered Cu oxides starting with 
hole-doped \ce{La2CuO4}\cite{Bednorz1986} followed soon thereafter by \ce{YBa2Cu3O7}\cite{Hor1987} that broke the 
all-important liquid \ce{N2} (77\,K) temperature barrier. These classes of materials are the only ones to date 
that have broken the liquid \ce{N2} temperature barrier without requiring high pressures. The layered oxide 
superconductors flagrantly violate the rules laid down by Matthias that low-dimensional materials, and specifically oxides
are bleak compositional spaces to seek superconductivity.\cite{Matthias1955} Recently, layered oxides of Cu have been 
joined in displaying superconductivity by some of their Ni counterparts.\cite{Li2019,Zhang2025}

The motivation behind this perspective is that both textbooks and the literature are rife with confounding statements  
in regard to what the essential ingredients are for a material to display metallic conduction. By focusing 
on the all-important oxides, we attempt to cast light on this topic. 
In virtually all textbooks, the erroneous assumption is often made that creating mixed valence and hence partial 
band filling through adding or removing electrons (respectively reducing or oxidizing certain ions) is sufficient 
to create metallic conduction. However, these suggestions ignore the need for the added charge carriers to be 
mobile: something that is crucially dictated by the bandwidth. 

As a notable example, the idea that ``LK-99''\cite{Lee2023}, an apatite phosphate of divalent lead with a formula close to 
\ce{Pb_{10}(PO4)6(OH)2}\cite{Griffin2023} can be doped to form a metal, and even one that is superconducting --- 
is cause for alarm. Phosphates and other oxyanion-rich compounds (borates, carbonates, silicates, sulfates, 
nitrates, phosphates, \textit{etc.}) have never been made metallic at ambient pressure because they lack the extended 
covalency that is required to create disperse bands capable of moving charge. 

Another domain where bandwidth underpins properties is the distinction between insulators and semiconductors, which is often 
attributed to the magnitude of the band gap.\cite{Callister2020,Woodward2021} However, this is questionabbe. For example, 
rutile \ce{TiO2} which is widely regarded as an insulator, has an experimental band gap of 3\,eV \cite{Pascual1978,Amtout1995} 
whereas \ce{Ga2O3}, which is an emerging wide band gap semiconductor,\cite{Higashiwaki2024} has a larger bandgap, close 
to 5\,eV.\cite{Orita2000} The bandwidth of conducting states in \ce{Ga2O3} is clearly necessary for this distinction since 
excited or doped electrons/holes are required to be mobile for a material to be considered a semiconductor. Related to this 
distinction is that that optical absorption edges in narrow-band materials are frequently confounded with absorption across a band 
gap.\cite{Morgan2023,Morgan2024} 

A third example of the role of bandwidth being largely ignored is manifested when attempts are made to dope mobile carriers 
into the insulating frustrated magnetic materials that have been proposed as candidates for quantum spin-liquid 
ground states.\cite{Kelly2016} A related misunderstanding is the conflation of flat bands that arise in the electronic structure
due to orbital frustration --- in kagome nets for example --- with ``molecular'' flat bands that arise due to the absence of 
extended covalency. The former can lead to interesting physics. The latter would only ensure insulating behavior and is highly 
unlikely to be a ingredient for superconductivity.\cite{Stanev2018}

\begin{table}[h]
\small
\caption{Defining terms}\label{Table:terms}
\begin{tabular}{|p{1.6in}|p{2.2in}|p{2.2in}|}
\hline
\textbf{Term} &  \textbf{Definition} &   \textbf{Consequence} \\ 
\hline
Bandwidth/Dispersion & Energetic range of electronic band ($E_{max} - E_{min}$) & Large bandwidth = carrier delocalization/ Small bandwidth = carrier localization\\
\hline
Hybridization        & Mixing of orbitals on same atom                          & Geometry and structure \\     
\hline
Localized covalency  &  Bonding within discrete entities like oxyanions (PO$_4^{3-}$, CO$_3^{2-}$, \textit{etc.})
                                                                                & Localized electrons and dispersionless (flat) bands \\
\hline
Extended covalency   &  Covalent bonding along 1D, 2D, or 3D frameworks (such as the 3D Re$^{6+}$--O$^{2-}$--Re$^{6+}$ interactions in \ce{ReO3})        
                                                                                & Dispersive bands and mobile electrons\\                                                             
\hline
+/$-$ Inductive effects &  Donation (+) or removal ($-$) of electron density by spectator ions
                                                                                & Electro(negative/positive) spectator ions 
                                                                                destabilize/stabilize high oxidation states (\textit{eg.} NiO has Ni$^{2+}$
                                                                                \textit{vs.} LaNiO$_3$ has Ni$^{3+}$)\\
\hline
Fajans' rules & Greater cationic charge and smaller cation size contribute to winning back anion charge density & High cation oxidation states increase covalency and can therefore impact bandwidth \\
\hline 
Nature of d orbitals & The order of covalency is usually 3d $<$ 4d $\approx$ 5d & 4d and 5d oxides are more likely to be metallic\\
\hline         
\end{tabular}
\normalsize
\end{table}

Throughout this perspective, we interchangeably use the terms bandwidth and band dispersion. In the interest of brevity, 
we present some key terms in Table\,\ref{Table:terms} to help familiarize readers with the language that we will employ
to describe electronic structure features and some of the structural and compositional features they originate from. 

The approach here, based around the details of crystal chemistry, is contrasted with, but is not in conflict with 
established ideas for when metallic behavior is observed, such as the landmark work of Zaanen, Sawatzky, and 
Allen\cite{Zaanen1985} who mapped compounds onto the space of correlation $U$ and anion-to-cation charge-transfer 
energies $\Delta$, scaled by the extent of hopping $t$ (related to the bandwidth $W$), finding metallic behavior 
only when $U/t$ or $\Delta/t$ are below a certain threshold.

The oxides described here are ones that display metallic behavior or can be driven metallic by small perturbations 
of temperature or composition, with a few exceptions presented for conceptual comparison.  For the purpose of this contribution, 
we eschew the formal definition of a metal as a crystal with a Fermi surface\cite{Kittel2018} and employ instead working definitions 
based around concept of the maximum metallic resistivity\cite{Ioffe1960,Mott1972} close to 10$^{-2}$\,$\Omega$\,cm. 
Materials with room-temperature resistivities that are smaller than this value, and conversely, materials with conductivities 
that are larger than around 10$^2$\,S\,cm$^{-1}$ are considered metals, and they would display a positive 
temperature-coefficient of resistivity (TCR). A positive TCR means resistivities increase (and conductivities 
decrease) upon heating, which is the opposite of any activated process. 

We employ first-principles density functional theory (DFT) calculations to create electronic structure depictions 
comprising the band structure, density of states (DOS), and crystal orbital Hamilton populations (COHP) in the 
region of the Fermi energy. The electronic structures presented here are not meant to be authoritative, and most, 
if not all of the oxides described here have been studied by several others in much greater detail, at various 
levels of sophistication. We employ the widely used PBE functional\cite{Perdew1996}, and place trust in the 
trends that are obtained, even if absolute energetics and more accurate band structures require other functionals, 
corrections, or beyond-DFT approaches. Most of the electronic structures presented here can also be found on the 
Materials Project.\cite{Jain2013} 

From these electronic structures, the simplest takeaway is that for metallic behavior, a bandwidth close to 
2\,eV appears to be an essential feature of the well-known metallic oxides. It is unlikely, in our 
estimation, that any material with a bandwidth that is smaller than 1\,eV will ever become metallic.

\section{The oxides}\label{sec2}

\begin{figure}
    \centering
    \includegraphics[width=1\textwidth]{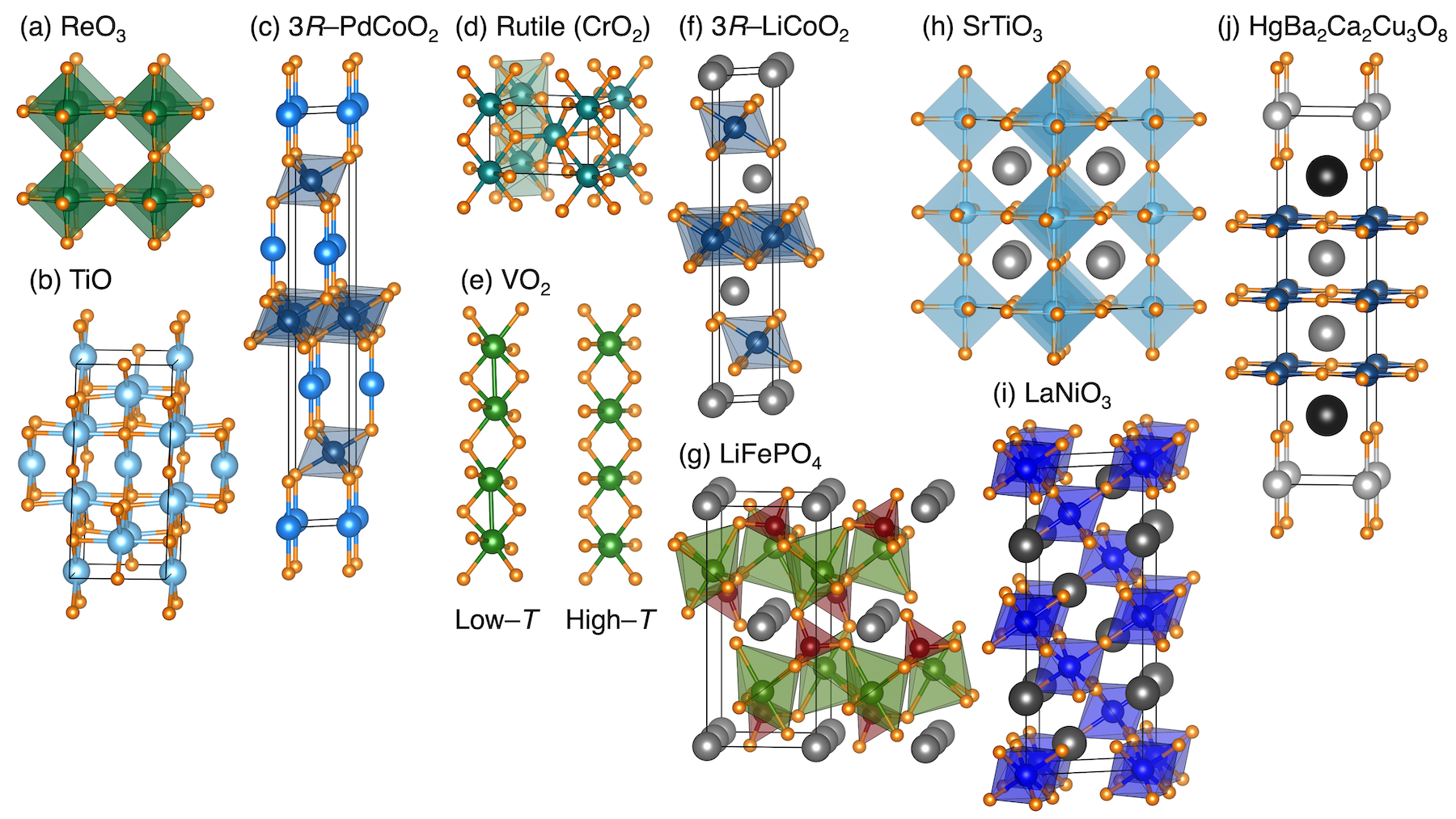}
    \caption{Crystal structures of the different compounds described in this perspective. (a) Cubic \ce{ReO3}, (b) 
    the defect rock-salt structure of TiO, (c) the layered delafossite $3R$--\ce{PdCoO2}, (d) the rutile structure of
    \ce{CrO2}, (e) low-- and high--$T$ variants of (part) of the structure of rutile \ce{VO2} displaying metal--metal 
    bonding in the low--$T$ state, (f) the layer-ordered rock-salt structure of $3R$--\ce{LiCoO2}, (g) the structure of 
    the olivine phosphate \ce{LiFePO4}, (h) the tetragonal ground state structure of perovskite \ce{SrTiO3},
    (i) the rhombohedral structure of perovskite \ce{LaNiO3}, and (j) the triple-\ce{CuO2}-layered, tetragonal structure 
    of \ce{HgBa2Cu2Cu3O8} superconductor parent.}
    \label{Fig:Structures}
\end{figure}

Crystal structures of the different oxide compounds whose electronic structures are detailed in this work are 
depicted in Figure\,\ref{Fig:Structures}. These range from simple oxides (binary compounds) like \ce{ReO3} to 
ternary compounds like \ce{SrTiO3}, quaternary \ce{LiFePO4} and the quinary superconductor parent compound 
\ce{HgBa2Ca2Cu3O8}. Most of these oxides are known to be metallic or can readily be doped into 
a metallic state (for example, \ce{SrTiO3} and \ce{LiCoO2}). It is seen that in all of the metallic (or 
proximal-to-metallic) oxides, the sites on which charge carriers reside are never far apart. The other 
observation is that in all of the compounds, the connectivity between the transition metal sites and the 
intervening oxygen is always at least 2 or 3-dimensional. The only example of an oxyanion compound that 
is presented here for completion is the olivine phosphate \ce{LiFePO4}. Oxyanions perforce increase the 
separation between metal centers and impact (reduce) the extended metal--O--metal or connectivity that is found in 
simple oxides.

\begin{figure}
    \centering
    \includegraphics[width=1\textwidth]{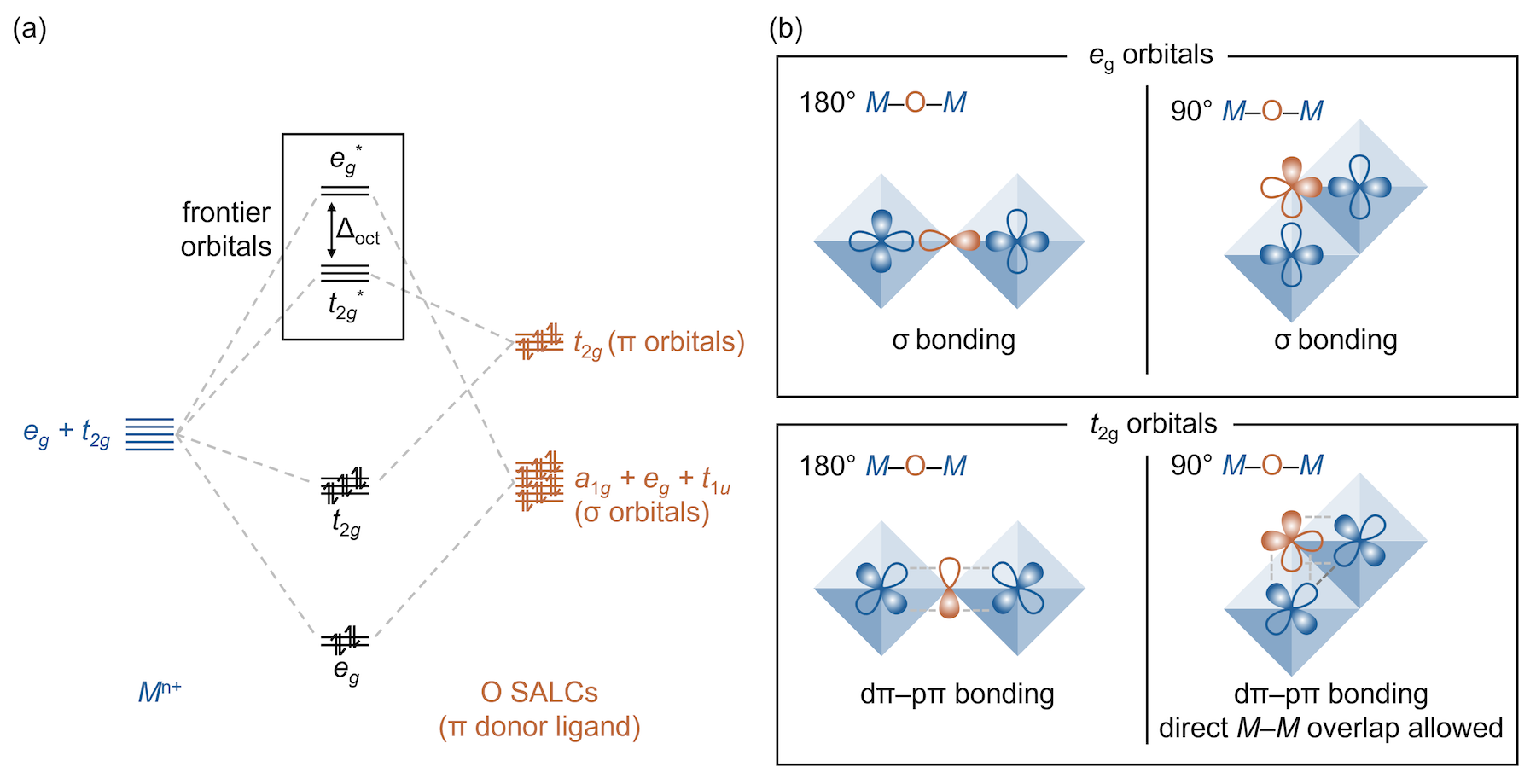}
    \caption{Local bonding environments within $M$O$_6$ octahedra. (a) Partial molecular orbital diagram highlighting antibonding 
            frontier orbitals given occupation of $M$ d orbitals (adapted from \cite{Housecroft2012}). (b) Orbital configurations 
            for $t_{2g}$ and $e_g$ filling based on corner-sharing or edge-sharing $M$O$_6$ octahedra.}
    \label{Fig:bonding}
\end{figure}

Many of the oxides presented here contain $M$O$_6$ octahedra, the molecular orbital (MO) diagram of which is displayed
in Figure\, \ref{Fig:bonding}. Since O is a $\pi$-donor ligand with filled symmetry-adapted linear combinations (SALCs), 
the filling of $t_{2g}^*$ and $e_g^*$ frontier MOs depends on the $M$ d orbital filling.\cite{Woodward2021,Housecroft2012} 
This means that as long as $M$ has above a d$^0$ configuration, the Fermi level interactions will be antibonding. Crystal 
orbital Hamilton populations (COHPs)\cite{Dronskowski1993} provide this bonding information by weighting the density of states by hopping between
orbital pairs. A negative COHP value describes a bonding interaction since energy is \textit{lowered}. It is the convention
convention to plot $-$COHP such that bonding interactions are shown as positive values and antibonding are negative.\cite{Dronskowski1993}
Integrating the $-$COHP along the energy axis yields the energy of the interaction,\cite{Dronskowski1993}  but as we are considering
widely varying families of oxides, we employ the COHP in only a qualitative fashion.

\subsection{d$^1$ ReO$_3$ --- the archetype}

\begin{figure}
    \centering
    \includegraphics[width=0.5\textwidth]{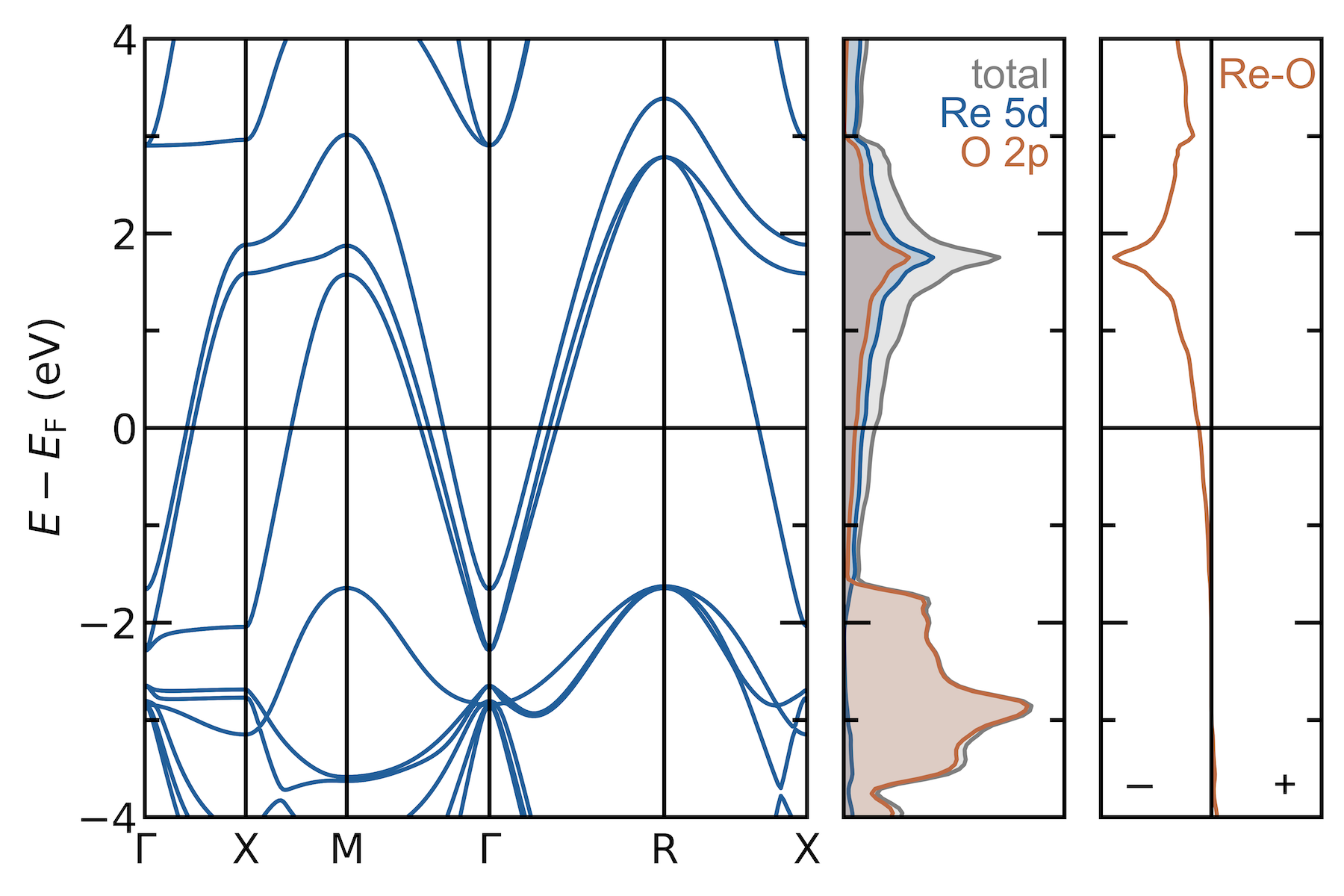}
    \caption{Electronic structure of \ce{ReO3} displaying Fermi-level states with bandwidths  $W$ = 5\,eV. The DOS and $-$COHP
            indicate that these states are covalent Re 5d and O 2p states with antibonding character as expected for materials 
            with $M$\ce{O6} octahedra and partial filling of the d orbitals.}
    \label{Fig:ReO3}
\end{figure}

\ce{ReO3} is one of the best known oxide metals with low temperature resistivities only slightly larger than 
the best electrical conductors like Cu and Ag. The crystal structure is simple, with fully connected and extended 180$^{\circ}$ 
interactions in all cardinal directions. The electronic structure is displayed in Fig\,\ref{Fig:ReO3} and shows a 
disperse and linear bands with width $W$ = 5\,eV. The structure does not permit direct metal--metal orbital overlap 
between Re d and O p orbitals as seen from the $-$COHP which displays antibonding at $E_F$. This interaction is of d$\pi$--p$\pi$ character 
because of the $t_{2g}^1$ crystal field configuration as depicted in Figure\,\ref{Fig:bonding}. Orbital interactions with $\pi$ character 
do not normally give rise to high covalency and disperse bands but the combination of the highly oxidized Re$^{6+}$ state (see Fajans' Rules
in Table\,\ref{Table:terms}), the more  extended 5d orbitals of Re, and the 3D framework of corner-sharing \ce{ReO6} octahedra lead 
to strong extended covalency in \ce{ReO3} as reflected in the remarkable bandwith. 

\subsection{d$^2$ TiO and d$^9$ PdCoO$_2$ --- direct metal--metal overlap and metallic behavior}

\begin{figure}
    \centering
    \includegraphics[width=0.5\textwidth]{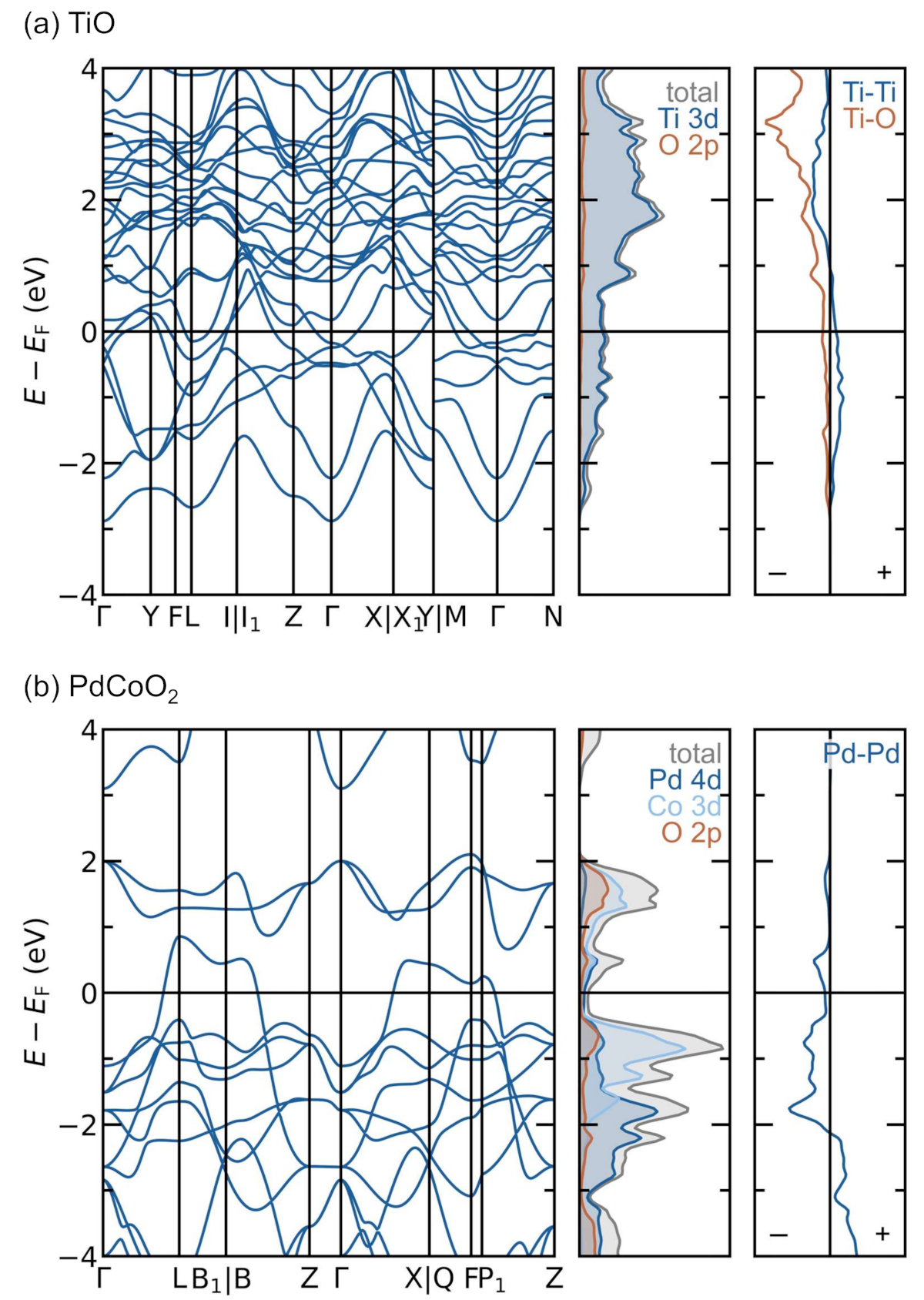}
    \caption{Electronic structures of (a) \ce{TiO} and (b) \ce{PdCoO2} showing disperse bands and states at the 
    Fermi energy associated with direct metal--metal interactions.}
    \label{Fig:TiO-PdCoO2}
\end{figure}

The simplest oxide metals could be rock-salts monoxides \textit{M}O, but these formulations are rarely 
as simple as they seem. While later transition metal oxides (MnO, CoO, NiO) are correlated, magnetic, 
insulators,\cite{Imada1998} early transition metals form metallic oxides. However, they often adopt
complex superstructures in order to accommodate metal-metal bonding, which is driven by two factors: the more
extended nature of d orbitals on early transition metals, and the edge-sharing of octahedra in the rock-salt 
aristotype (the reigning structure type from which ordered variants are derived) that allows the metals to approach one 
another and even form a bond. This is also important for rutile compounds as we will describe when we discuss \ce{VO2}. 
Examples of complex rock-salt derived superstructures are seen in the structures of TiO and NbO.\cite{Burdett1984}

Figure\,\ref{Fig:TiO-PdCoO2}(a) displays the electronic structure of monoclinic TiO that displays several short Ti--Ti
contacts (less than 2.8\,\AA\/ in distance) and this gives rise the direct metal--metal interaction seen in the 
$-$COHP, contributing to electrical conductivity and the silvery lustre of this oxide. The electronic structure of 
\ce{PdCoO2} is displayed in Figure\,\ref{Fig:TiO-PdCoO2}(b). \ce{Pd\textit{M}O2} and \ce{Pt\textit{M}O2}
delafossite oxides have long been known to possess very short contacts between metals in the Pd/Pt plane in 
addition to demonstrating an unusual, monovalent and nonmagnetic d$^9$ state on the 
Pd/Pt.\cite{Shannon1971,Seshadri1998} In this compound, d$^6$ Co$^{3+}$ is low-spin, diamagnetic, and has a filled 
$t_{2g}^6$ state which does not contribute to the electrical transport. In the delafossite, the linear dispersion of the 
bands crossing the Fermi level is notable, and potentially contributing to the very low effective masses of charge 
carriers and thereby, the exceptional charge transport behavior.\cite{Mackenzie2017}

\subsection{The rutiles: d$^0$ TiO$_2$, d$^2$ CrO$_2$, d$^4$ RuO$_2$, d$^{10}$ SnO$_2$}

\begin{figure}
    \centering
    \includegraphics[width=1\textwidth]{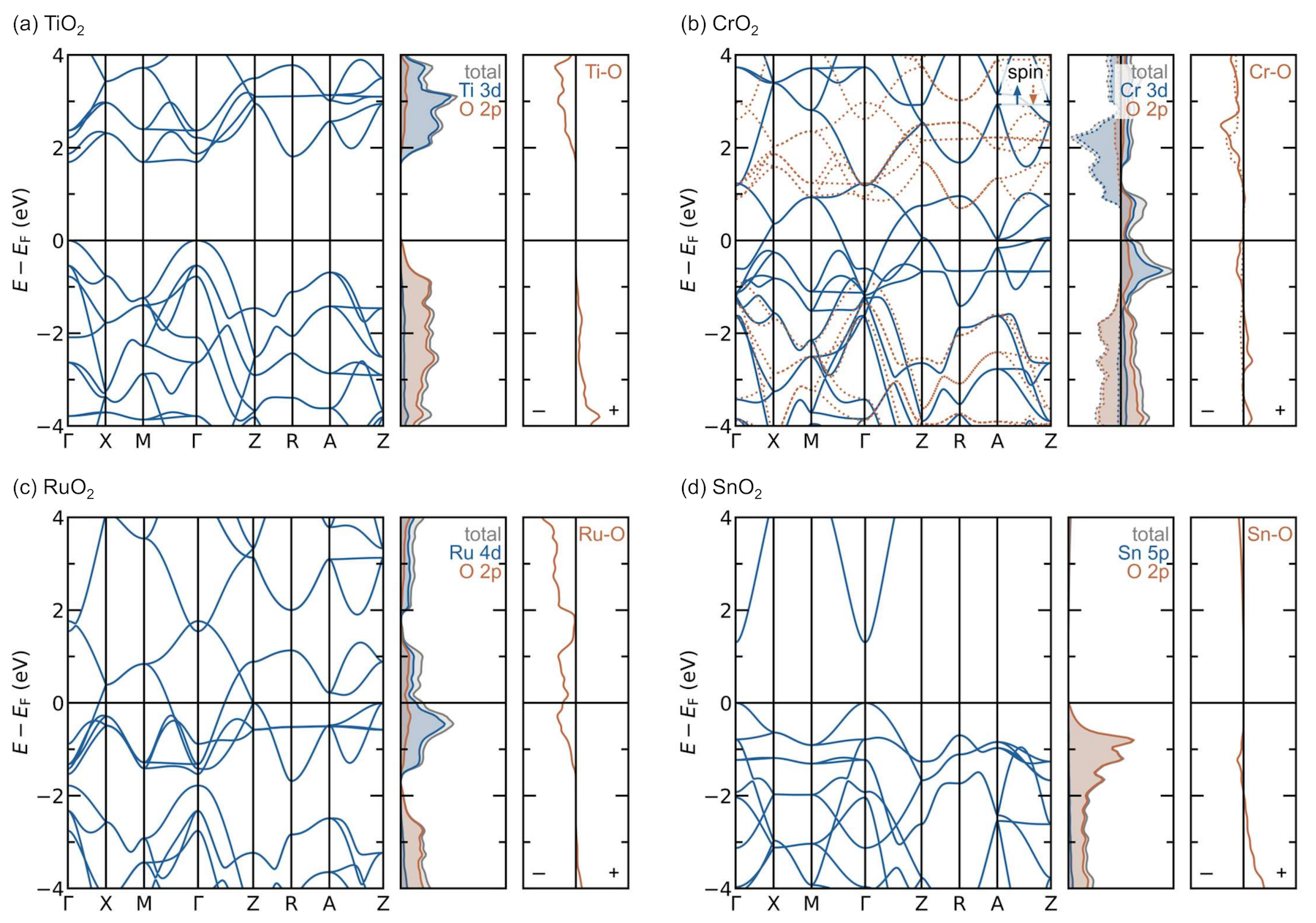}
    \caption{Electronic structures of selected rutile materials highlighting a variety of conduction behavior. 
            (a) \ce{TiO2} is a band insulator, (b) \ce{CrO2} is a ferromagnetic half-metal, (c) \ce{RuO2} is 
            a metal, and (d) \ce{SnO2} is a semiconductor. With a completely empty or full d shell, \ce{TiO2} 
            and \ce{SnO2} show O 2p valence states and no $M$--O interactions at the Fermi level. \ce{CrO2} and 
            \ce{RuO2} possess partially filled d-orbitals and, therefore, have mixed $M$ d and O p states at 
            the Fermi level with antibonding interactions indicated in the $-$COHP.}
    \label{Fig:rutiles}
\end{figure}

The rutile structural class is a relevant family to discuss since it displays the diversity of electronic 
conduction behavior among oxides. Figure\,\ref{Fig:rutiles} details the electronic structures of selected
rutile materials spanning from insulators to metals. \ce{TiO2} is a d$^0$ insulator with an experimental band gap 
around 3 eV.\cite{Pascual1978,Amtout1995} \ce{CrO2} and \ce{RuO2} possess partially filled d orbitals (d$^2$ and d$^4$) 
and are a ferromagnetic half-metal and metal, respectively.\cite{Schwarz1986,Korotin1998,Glassford1993,Smolyanyuk2024} 
\ce{SnO2} with fully filled d orbitals is similarly gapped like \ce{TiO2}, however, this material is an n-type semiconductor 
becoming metallic when $E_F$ is raised into the dispersive conduction states through electron doping.
\cite{Batzill2005,Kilic2002} The comparison between \ce{TiO2} and \ce{SnO2} shows the success of using this approach 
of integrating bandwidth into understanding electronic conduction. While standard DFT parameters yield comparable 
band gaps for both materials, the conduction bandwidth of \ce{SnO2} is significantly greater than that of \ce{TiO2}. 
When electrons are excited into the conduction states, the electrons will be more mobile in \ce{SnO2} than in \ce{TiO2}, 
a necessary condition for a semiconductor. 

\subsection{d$^1$ VO$_2$ -- a band insulator?} 

\begin{figure}
    \centering
    \includegraphics[width=0.5\textwidth]{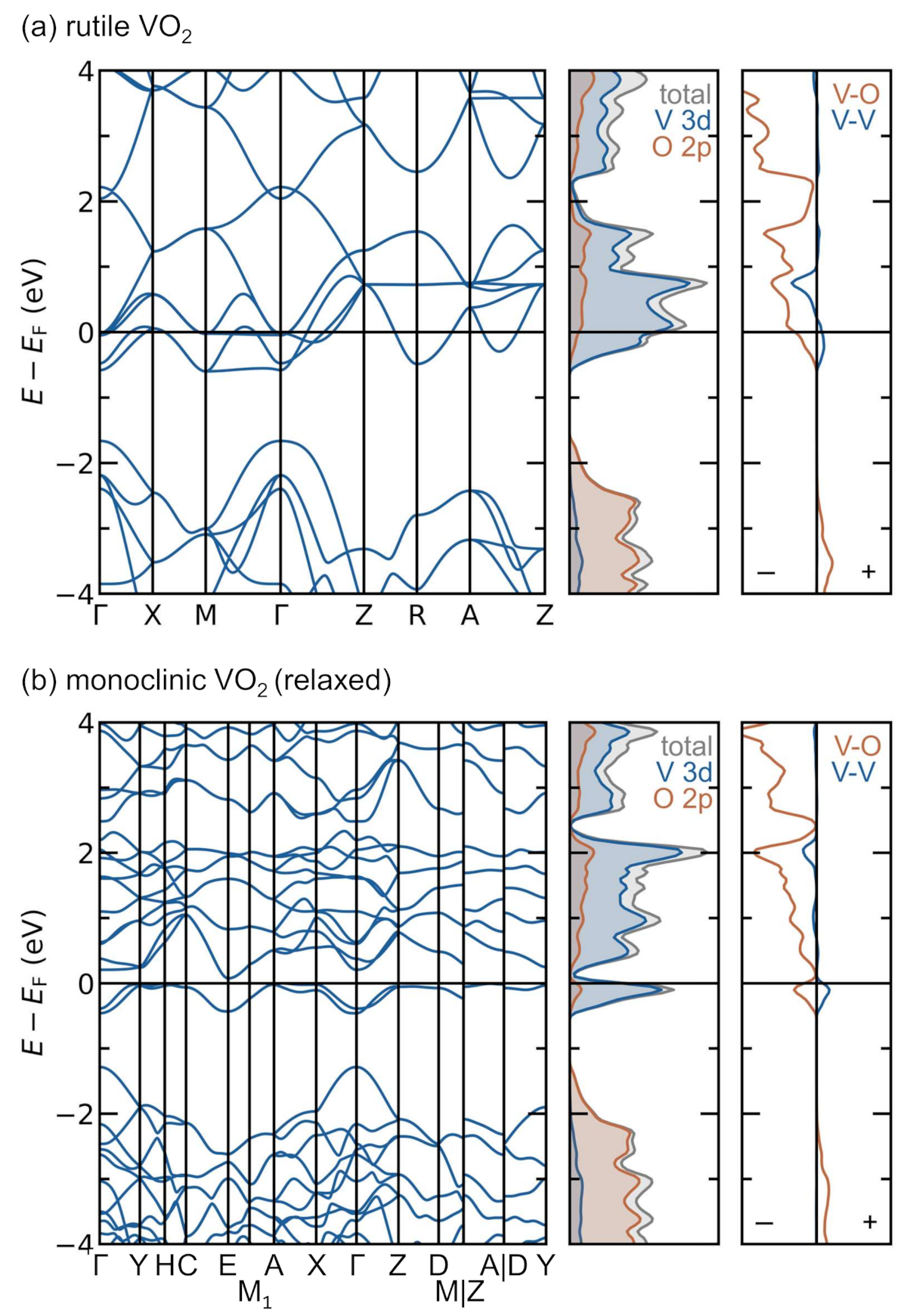}
    \caption{Electronic structures of (a) high-temperature rutile \ce{VO2} and (b) low-temperature monoclinic \ce{VO2}. 
            A small gap opens in the electronic structure of the low-temperature phase stemming from V--V bonding and
            antibonding separating, as seen in the $-$COHP.}
    \label{Fig:VO2}
\end{figure}

Another member of the rutile family at high temperatures, \ce{VO2} undergoes a metal--insulator transition
accompanied by a structural transition to a monoclinic unit cell (space group $P2_1/c$) around 
340\,K.\cite{Andersson1954,Andersson1956} Since the transitions occur simultaneously, the origin of the transition, 
whether due to a structural, electronic, or combined instability, has been the source of controversy.\cite{Eyert2002} 
Studying the local interactions around the transition can provide insight into the driving force. The edge-sharing octahedra 
within the rutile structure of \ce{VO2} allow for close proximity of neighboring V atoms which dimerize in the 
low-temperature structure.\cite{Corr2010} The V--V bonding upon dimerization from the high-temperature to 
low-temperature structure is seen in the $-$COHP in Figure\,\ref{Fig:VO2}. Pairs of the d$^1$ electrons on neighboring 
V$^{4+}$ form localized bonding and antibonding molecular orbitals akin to what is seen in the H$_2$ molecule, with bonding regions 
below the Fermi energy narrowing, becoming more prominent, and splitting off from antibonding regions above the Fermi energy. 
This locks up the d$^1$ conduction electrons in the metal--metal bonds, meaning that \ce{VO2} at low temperatures 
effectively becomes a band insulator where all electrons are accounted for within bonds.\cite{Hiroi2015}  

\subsection{Cathode materials: d$^6$ LiCoO$_2$ and d$^6$ LiFePO$_4$ } 

\begin{figure}
    \centering
    \includegraphics[width=0.5\textwidth]{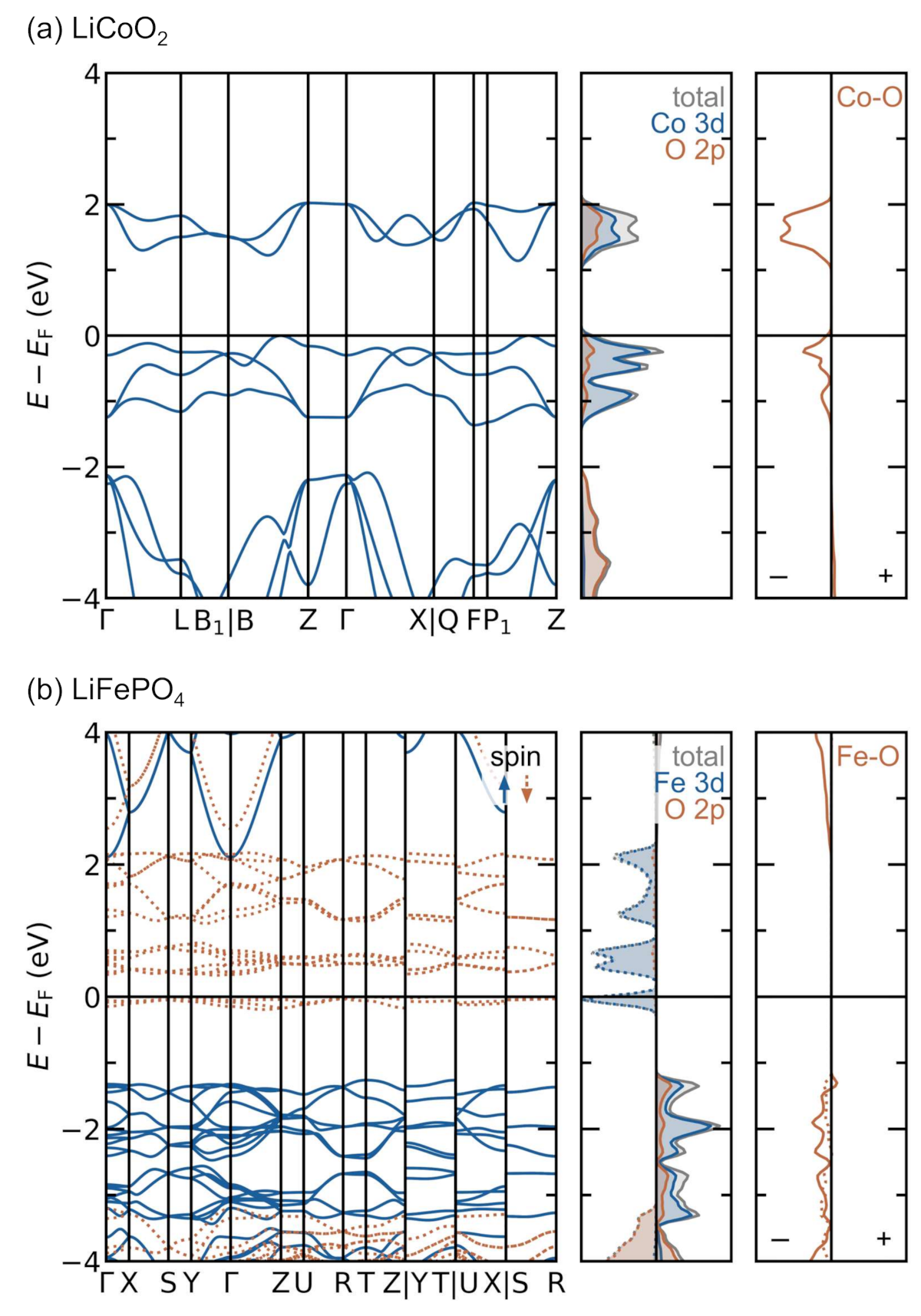}
    \caption{Electronic structures of (a) \ce{LiCoO2} and (b) \ce{LiFePO4}. The wider valence bands in \ce{LiCoO2} 
            facilitate metallic conduction upon hole doping though delithiation.}
    \label{Fig:cathodes}
\end{figure}

The prototypical cathode material \ce{LiCoO2} features planes of edge-sharing \ce{CoO6} octahedra separated by Li
ion layers. With d$^6$ Co$^{3+}$ in a low-spin $t_{2g}^6$ state, the compound is a diamagnetic insulator, but 
becomes metallic under Li-deintercalation (hole-doping) conditions creating a partially filled band. The subsequent 
Co oxidation from +3 to +4 additionally favors greater Co--O covalency which aids metallic behavior. \ce{LiCoO2} 
is an example of a 90$^{\circ}$ $t_\mathrm{{2g}}$ system depicted in Figure\,\ref{Fig:bonding}(b) where the geometry permits $\pi$ 
overlap between $t_\mathrm{{2g}}$ orbitals via the oxygens, enabling electronic delocalization and metallic behavior. 
In contrast, the related compound d$^7$ \ce{LiNiO2} exhibits a 90$^{\circ}$ $e_{g}$ configuration (Figure\,\ref{Fig:bonding}(b)), 
where the $\sigma$-bonding between $e_{g}$ orbitals is diminished compared to the 180$^{\circ}$ case with corner-sharing 
octahedra, resulting in suppressed overlap and non-metallic behavior. Another material popularized as a cost-effective and nontoxic 
alternative to the transition metal oxide based cathodes (\ce{LiNiO2}, \ce{LiMnO2}, \ce{LiCoO2}, or the combined 
``NMC" \ce(LiNi$_x$Mn$_y$Co$_{1-x-y}$O$_2$)) is the polyanion oxide with the olivine crystal structure, \ce{LiFePO4}. 
This material is insulating with a suggested experimental band gap --- always challenging to define in a correlated 
compound with magnetism --- of around 4\,eV \cite{Zaghib2007} and requires additional processing, notably preparing as 
small particles that are conducting carbon-coated, for application as a battery electrode. The localized 
covalency of the phosphate groups and the low connectivity throughout the structure results in flat bands indicative 
of trivial carrier localization in the electronic structure in Figure\,\ref{Fig:cathodes}(b), supporting the idea that efforts 
to make this material intrinsically more conducting are futile.\cite{Chung2002,Herle2004} While modeling the 
experimental band gap requires additional treatment of electronic correlation with a Hubbard \textit{U} around 4 eV, 
comparing the bandwidths of the \ce{LiFePO4} conduction states and the \ce{LiCoO2} valence states under standard DFT 
conditions serves as a diagnostic test for metallicity.\cite{Zhou2004}

\subsection{Perovskites: d$^0$ SrTiO$_3$ and  d$^7$ LaNiO$_3$} 

\begin{figure}
    \centering
    \includegraphics[width=0.5\textwidth]{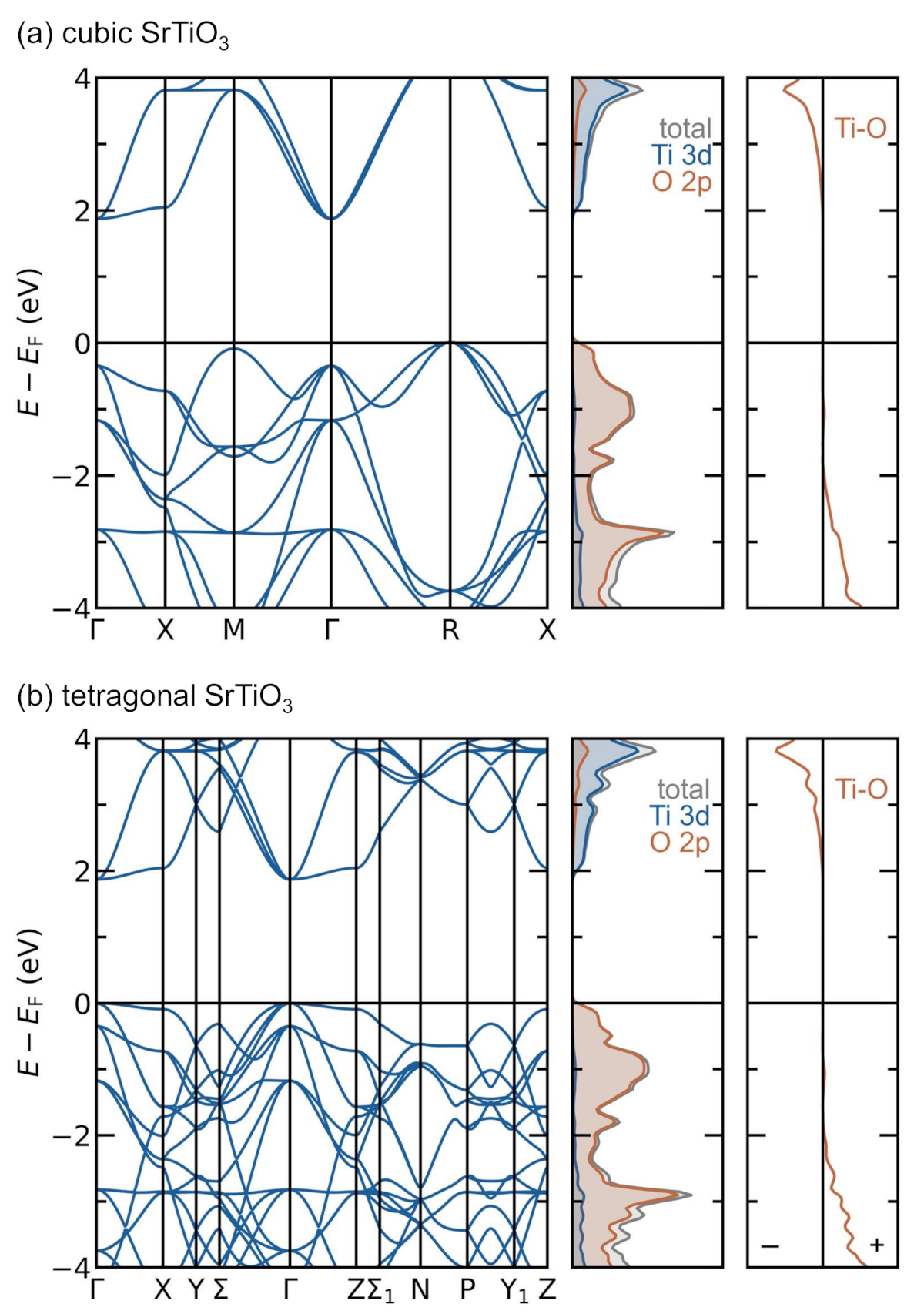}
    \caption{Electronic structures of (a) high-temperature cubic \ce{SrTiO3} and (b) low-temperature tetragonal \ce{SrTiO3}
            displaying wide conduction bands that support metallic behavior with electron doping.}
    \label{Fig:STO}
\end{figure}

\begin{figure}
    \centering
    \includegraphics[width=0.5\textwidth]{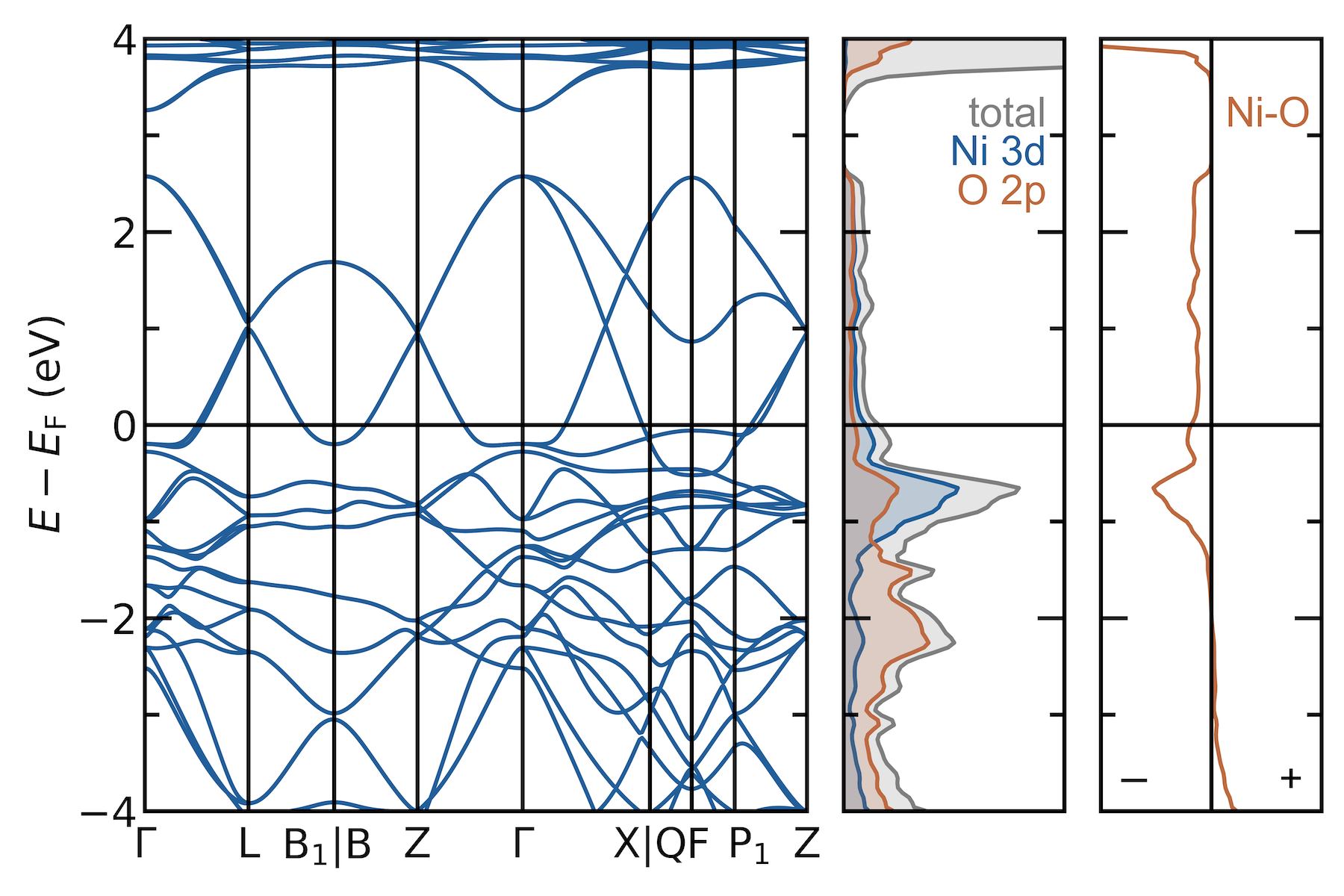}
    \caption{Electronic structure of \ce{LaNiO3} with dispersive bands at the Fermi level owing to partial $e_g$ occupation 
            and a Ni$^{3+}$ oxidation state.}
    \label{Fig:LaNiO3}
\end{figure}

Perovskite oxides with the general formula $AM$O$_3$ have a long history of functional versatility ranging from 
magnetism to ferroelectricity to superconductivity, \textit{etc.}\cite{Bhalla2000} The $A$ and $M$ sites can accommodate 
many elements enabling the tunability of crystal chemistries and physical properties. The structure of \ce{ReO3} 
is a good proxy for understanding the interactions within perovskites.\cite{Evans2020} Similar to \ce{ReO3}, 
the perovskites feature a network of three-dimensionally corner-connected $M$O$_6$ octahedra where O is covalently 
bonded to the more electronegative, and usually charged $M$-site cation compared to the ionic interactions 
with the $A$-site cation. In the limit that the $A$-site cation donates all valence electrons to the $M$O$_6$ 
framework, the electronic structure of cubic perovskites matches that of \ce{ReO3} with the Fermi level dependent 
on the $M$ d orbital filling.\cite{Woodward2021} This resemblance of the band structure can be seen for cubic 
\ce{SrTiO3} in Figure\, \ref{Fig:STO}a. Since \ce{SrTiO3} has a 3d$^0$ configuration, the 
electronic structure has a gap between the O 2p valence states and the Ti 3d conduction states. Like 
many perovskites, \ce{SrTiO3} distorts from the cubic $Pm\overline{3}m$ structure to an antiferrodistortive 
tetragonal $I4/mcm$ phase with tilted \ce{TiO6} octahedra around $T$ = 105\,K.\cite{Muller1968,Shirane1969} The nature 
of the antiferrodistortive and ferroelectric instabilities have been extensively studied since \ce{SrTiO3} is an 
incipient ferroelectric (or quantum paraelectric) meaning that a polar transition is suppressed by quantum 
fluctuations.\cite{Sai2000,Aschauer2014,Zhu2024} The electronic structure of the low-temperature phase shown in 
Figure\,\ref{Fig:STO}b displays dispersive and linear conduction bands similar to the cubic phase along with 
saddle points near the Fermi level. The wide conduction bands are responsible for metallicity at very low electron 
doping concentrations since electrons from dopants are mobile within those states.\cite{Spinelli2010}

Rare-earth nickel oxide perovskites (formula \textit{Ln}\ce{NiO3}, \textit{Ln} = Y or Lanthanoid) are an interesting class of 
oxides from the perspective of electrical and magnetic behavior, displaying metal--insulator transitions 
controlled by the size of the rare-earth ion. Smaller radii facilitate greater \ce{NiO6} octahedral tilting resulting
in smaller Ni--O--Ni bond angles and therefore less disperse conduction bands.\cite{Zhou2014} The distortion has 
consequences on the metal--insulator ($T_{MIT}$ increases) and antiferromagnetic transitions ($T_N$ decreases) of these 
materials.\cite{Balachandran2013} With the largest rare earth cation, \ce{LaNiO3} deviates from this behavior, 
remaining metallic down to low temperatures with no magnetic ordering in fully oxygenated and stoichiometric 
samples,\cite{Zhang2017} but with evidence for proximity to an antiferromagnetic quantum critical point.\cite{Liu2020}
The relatively high oxidation state (+3) on a late transition metal like Ni in \ce{LaNiO3} and positive inductive effects
from La$^{3+}$ contribute to covalent Ni--O bonding (see Inductive Effects and Fajans' rules in Table\,\ref{Table:terms}) 
leading to wide bands shown in Figure\,\ref{Fig:LaNiO3}. The $t_{2g}^6e_g^1$ configuration also helps create dispersive Fermi-level 
states since the $\sigma$ bonding associated with the $e_g$ filling has more overlap than $\pi$ interactions. Clearly,
the computed bandwidth of \ce{LaNiO3} is at the lower limit for metallicity since it is the only member of the 
\textit{Ln}\ce{NiO3} to possess a metallic ground state.

\subsection{Copper oxide superconductors --- the example of d$^9$ HgBa$_2$Ca$_2$Cu$_3$O$_8$}

\begin{figure}
    \centering
    \includegraphics[width=0.5\textwidth]{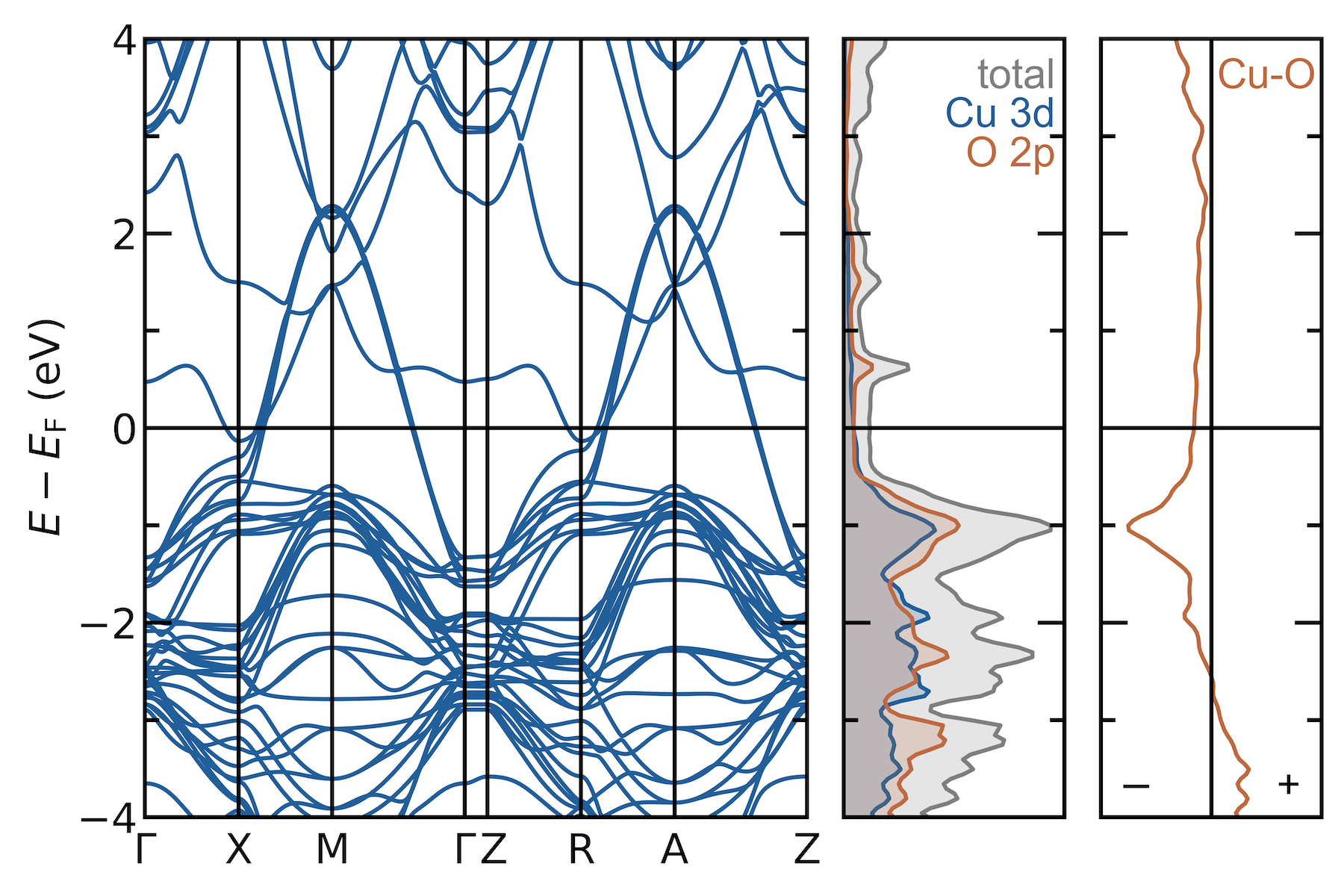}
    \caption{Electronic structure of \ce{HgBa2CaCu3O8} featuring linear and dispersive bands at the Fermi level indicative
            of high carrier mobility. The high dispersion speaks to the expected strong antiferromagnetic coupling in the proximal, 
            insulating ground state.}
    \label{Fig:HgBa2Ca2Cu3O8}
\end{figure}

Complex layered copper oxides are currently the highest temperature superconductors at ambient pressure, 
and it is generally agreed that spin-fluctuation mediated pairing plays a key role in driving 
superconductivity.\cite{Mazin2010,Scalapino2012} These systems share the common structural feature of Cu--O 
planes that carry the supercurrent, separated by charge reservoir layers.\cite{Keimer2015} The layered 
superconducting copper oxides have perovskite-derived structures similar to Ruddlesden-Popper compounds 
with two-dimensional \ce{CuO2} networks that are covalent due to the $\sigma$ interactions between 
Cu 3d$_{x^2-y^2}$ and O 2p$_\sigma$ orbitals.\cite{Rao1989_cuprates,Park1995} 
In Figure\,\ref{Fig:HgBa2Ca2Cu3O8}, we display the electronic structure of the triple-layered copper oxide 
superconductor \ce{HgBa2Ca2Cu3O8}. This compound was specially selected since it is one of the few (alloy) 
disorder-free superconductors, and also exhibits the highest superconducting transition temperature of 
133\,K.\cite{Schilling1993} If a 2+ charge were formally assigned to Hg, this compound should contain only d$^9$ Cu 
with a single hole in the d manifold, which would normally result in an antiferromagnetic insulator. Such a ground state is 
well established for d$^9$ \ce{La2CuO4}, the parent compound of all Cu oxide superconductors\cite{Vaknin1987} and 
it is well-known that standard DFT techniques struggle to capture this ground state.\cite{Lane2018} The overbinding 
flaws of DFT ensure here that the calculated electronic structure in Figure\,\ref{Fig:HgBa2Ca2Cu3O8} resembles that of a 
heavily hole-doped compound, and suffices for the purpose of this discussion. The electronic structure highlights 
the dispersive bands at the Fermi level with bandwidths of almost 4\,eV signifying high carrier mobility. 

To our knowledge, this high dispersion is almost a record among oxides, and there are several contributing factors.
The first is that transition metal d--levels are more stabilized (approaching closer to O p levels) as one traverses 
the row from Sc to Cu, making Cu the most covalent and most electronegative 3d transition metal. This stabilization of
d levels has familiar consequences including in developing the phase diagram of Zaanen, Sawatzky, and Allen,\cite{Zaanen1985} 
the ``redox competition'' scenario of Rouxel,\cite{Rouxel1996} and the energetics of electrode materials in 
Li--ion batteries,\cite{Hayner2012}. A second reason for the dispersion is that the structure type adopted by the copper 
oxide superconductors possess extended 180$^\circ$ Cu--O--Cu $\sigma$ interactions in both $x$ and $y$ directions, 
with half the $e_g$ orbitals oriented in a manner that permits strong covalency with oxygen p orbitals. 
Finally, ions such as Ba$^{2+}$ and Ca$^{2+}$ serve to further push the Cu d and O p states closer together by inductive 
effects,\cite{Misch2014} again enhancing the covalent nature of the interaction, giving rise to broad bands. The 
extended covalency in these compounds also contributes to strong antiferromagnetic superexchange in the undoped compounds. 
Strong antiferromagnetism in a layered, covalent compound appears to be a ingredients for high-temperature superconductivity 
and underpin why the copper oxides are special. It is an open challenge of whether there exist other 
structures/compositions that capture similar chemistry. While the electronic structure is somewhat 
reminiscent of what was presented for \ce{ReO3}, there are notably more features such as saddle points at $M$ and $R$ 
in the electronic structure of the superconductor, a consequence of the lower dimensionality of the crystal
and electronic structure,\cite{Friend1979} which are absent in the much ``cleaner'' electronic structure of cubic \ce{ReO3}.

\section{Flat bands: atomic \textit{vs.} topological}

\begin{figure}
    \centering
    \includegraphics[width=1\textwidth]{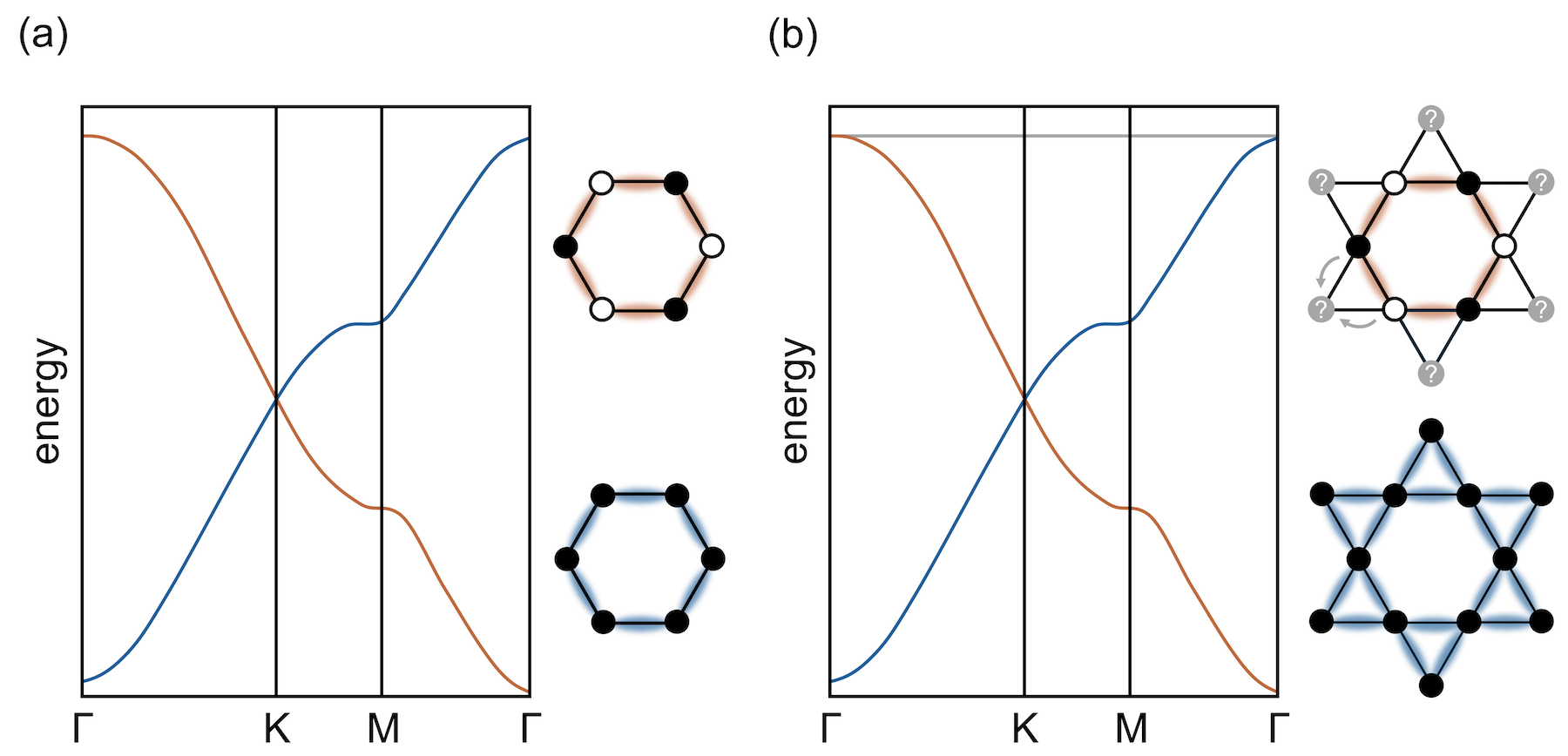}
    \caption{Tight-binding band structures of (a) honeycomb and (b) kagome networks with crystal orbital diagrams. In
            a honeycomb network, bonding (blue bonds) and antibonding (orange bonds) configurations with a two-atom basis
            result in two dispersive and crossing bands. The resemblance of the honeycomb and kagome atomic geometries leads 
            to similarities in the band structures. In the kagome structure, the same dispersive bands originate 
            from bonding and antibonding interactions within the hexagons. Since a kagome network has a three-atom basis,
            the orbital frustration in the triangles creates a flat band at high energy (as expected for antibonding). 
            Further explanations of these band structures can be found in \cite{Johnston1990,Jovanovic2022}. }
    \label{Fig:kagome}
\end{figure}

\begin{table}[h]
\caption{Comparison of atomic and topological flat bands}\label{Table:flat_bands}
\begin{tabular}{|p{2.5in}|p{2.5in}|}
\hline
\textbf{Atomic flat bands}                            & \textbf{Topological flat bands} \\
\hline
 Localized orbitals or separated atoms                & Extended wavefunctions  \\
\hline
Vanishing overlap between atomic wavefunctions        & Large orbital overlap and hopping between wavefunctions  \\
\hline
Minimal overlap suppresses electronic kinetic energy  & Quantum interference effects quench electronic kinetic energy \\
\hline
Systems with separated atoms or poorly overlapping orbitals (\textit{i.e.} localized 3d, 4f, or 5f orbitals) 
                                                      & Systems with nontrivial quantum geometry  \\
\hline          
\end{tabular}
\end{table}

In this discussion of bandwidth, it is useful to distinguish between bands with little to no dispersion. Flat or 
narrow bands, in which the carrier kinetic energy is quenched and many-body interactions dominate, can be electronic 
structure landmarks of emergent correlated phenomena like unconventional magnetism or high-temperature superconductivity.
These bands can be broadly categorized into two classes: i) atomic and ii) topological and this distinction becomes 
important when claiming properties arising from flat bands.\cite{Regnault2022} Atomic flat bands arise in systems with 
atomically-localized orbitals either due to the structure (molecular crystals, layered or low-dimensional materials, 
supercells \ldots) or composition (f-electron or localized d-electron elements). Flat bands originating from molecular 
structures where atoms are separated will not lead to exotic correlated physics because electrons are trivially localized 
(no orbital overlap or hopping). Topological flat bands are found in systems where a frustrated network creates destructive 
interference between electron hopping paths. The network geometry dictates electronic localization despite the presence 
of orbital overlap and electron hopping.\cite{Kang2020} Materials with a kagome network are examples of topological 
flat band systems where the frustration from the triangular sublattice creates localization within the hexagons as 
depicted in Figure\,\ref{Fig:kagome}. A comparison of atomic and topological flat bands is detailed in 
Table\,\ref{Table:flat_bands}.\cite{Johnston1990,Jovanovic2022,Regnault2022,Checkelsky2024}

Identifying the origin of a narrow band within an electronic structure is critical in the context of predicting 
behavior associated with strong correlations like unconventional superconductivity. Flat bands inherently enhance the
effect of inter-electron interactions since kinetic energy, which normally dominates over the correlation energy scale,
is minimized. However, the existence of a flat band especially near the Fermi level does not immediately imply that the 
correlated state produces unusual physics. A recent example of this type of prediction was the proposed room-temperature
superconductor known as ``LK-99" based on a Cu-doped lead hydroxyapatite (\ce{Pb9Cu(PO4)_6(OH)_2}).\cite{Lee2023} The Cu 
dopant states manifest as very narrow bands residing at $E_F$ with bandwidth $W$\,=\,0.1\,eV.\cite{CabezasEscares2024,Lai2024,
Griffin2023,Si2023,Kurleto2023,Jiang2023,Yang2023} The electronic structure of the undoped material reveals that almost all 
of the bands are narrow which is consistent with the molecular-like crystal structure that features isolated atoms and phosphate 
groups. When Cu is doped into the system replacing Pb, the resulting flat band is a consequence of trivial electron localization 
and, therefore, not an indication of flat-band induced superconductivity.\cite{Jiang2023,Georgescu2025}

Referencing other bands within the electronic structure like this, in addition to looking at the crystal structure,
can be a helpful tool in determining the origin of a flat band. If all bands are flat or all bands within a specific
Brillouin zone direction are flat, the system is likely molecular or layered, respectively. For example, almost all
of the bands in Figure\,\ref{Fig:cathodes}b have narrow bandwidths matching the molecular-like structure of \ce{LiFePO4}. 
This method allows for fast identification of trivial atomic flat bands even at simpler levels of density functional
theory without additional treatment of electronic correlation.  

To access emergent correlated phenomena a flat band must (i) sit near the Fermi level and (ii) have balanced 
kinetic and correlation energy contributions.\cite{Paschen2021,Georgescu2025} If the band is too flat 
(correlation energy $>>$ kinetic energy; $U >> W$), there will be no emergent physics because there is no electronic 
communication necessary for collective behavior. If the band is too dispersive ($W >> U$), the kinetic energy will 
wash out any correlation effects. This interplay of electron localization and delocalization arises naturally in 
topological flat bands where the structural network promotes localization despite (and even because of) the presence of 
kinetically active electrons.

\section{Outlook}

Studying the crystal structures, local bonding environments, and electronic structures of a broad and diverse 
class of materials like oxides can highlight structure-property relationships that are more generally applicable 
to crystalline inorganic materials. Here, we discussed a selection of mostly metallic or proximally-metallic oxides
spanning many structure types and functional domains to draw attention to electronic bandwidth as an indicator of 
carrier mobility and, therefore, conduction behavior. There are several key take-aways from analyzing the electronic 
structures of these selected oxides:\\

\noindent \textbf{(i) Extended covalent interactions produce wide bands and metallicity.} Strong covalency between atoms in 
extended multi-dimensional frameworks throughout the structure promote mobile carriers. In the oxides studied here
at a relatively simple level of theory, we find metallic behavior typically requires a bandwidth $W > 1$\,eV.\\

\noindent \textbf{(ii) Achieving metallicity or superconductivity \textit{via} doping requires added electrons or holes to be mobile}. 
Tuning the Fermi level by electron- or hole-doping to yield partially-occupied bands is only part of the equation to create electronic
conduction. The added electrons and holes must also be mobile (\textit{i.e.} associated with wide bands).\\

\noindent \textbf{(iii) Band gap magnitude is an insufficient metric to distinguish between semiconductors and insulators.} For example, 
the insulator \ce{TiO2} has an experimental band gap of 3\,eV \cite{Pascual1978,Amtout1995} whereas the semiconductor \ce{Ga2O3} has a 
band gap close to 5\,eV.\cite{Orita2000} Band gap magnitude cannot account for the difference in conduction in these materials. 
Instead, the bandwidth of the conduction states must also be considered since excited or doped electrons must be mobile for the material to be 
a semiconductor. This can be seen for \ce{TiO2} and \ce{Ga2O3} where the latter has wider conduction bands than the former, even 
with simple functionals that do not reproduce the experimental gaps.\\

\noindent \textbf{(iv) Flat or narrow bands near the Fermi level are not always signatures of emergent correlated physics} and are usually 
detrimental to collective electronic states. While topological flat bands due to nontrivial quantum geometry lead to interesting phenomena, 
atomic flat bands arising from separated atoms or localized orbitals do not support high-temperature superconductivity, \textit{etc.} 
It is critical to determine the origin of a flat band by analyzing the crystal and electronic structure to prevent mischaracterization.

\section{Methods}\label{secA1}
	
Electronic structures were calculated using the Vienna Ab-initio Simulation Package (VASP) v5.4.4 with the PBE
functional\cite{Perdew1996}, a plane-wave energy cutoff of 500\,eV, and projector-augmented wave pseudopotentials 
recommended for VASP v5.4.\cite{Blochl1994_PAW,Kresse1999} \textit{k}-point meshes were automatically generated 
in VASP with the length parameter set to 50. All structurally-related materials share the same \textit{k}-meshes 
(for example all rutile structures presented here were calculated using an 11 $\times$ 11 $\times$ 17 
$\Gamma$-centered mesh.) Experimental geometries were employed for all materials except the low-temperature 
phase of \ce{VO2} to allow for V--V dimer formation as reported in Ref.\,\cite{Corr2010}. The relaxation had a 
force convergence criterion of 10$^{-5}$\,eV\,\AA$^{-1}$ and ionic positions, cell volumes, and cell shape were 
allowed to change via the conjugate gradient algorithm. Secondary self-consistent calculations were performed 
with double the number of default bands to ensure proper functioning of the COHP post-processing package 
(additionally symmetry and spin-orbit coupling were turned off for the same reason). All self-consistent and 
electronic structure calculations had an energy convergence better than 10$^{-6}$\,eV. Self-consistent and 
DOS calculations were performed using the tetrahedron method with Bl{\"o}chl corrections.\cite{Blochl1994} 
Spin-orbit coupling (SOC) was included in the self-consistent and band structure calculation for \ce{ReO3} 
(separate non-SOC calculations were performed for the DOS and $-$COHP plots in Figure\,\ref{Fig:ReO3} since SOC 
had minimal impact on the DOS and SOC is incompatible with the COHP package). Spin-polarized calculations 
were performed with initialized magnetic moments for Cr in \ce{CrO2} and Fe in \ce{LiFePO4}. The AFLOW 
online tool was used to generate \textit{k}-point paths for the band structures calculations with the 
density per path set to 50 \textit{k}-points.\cite{Curtarolo2012} Band structures were plotted using the 
SUMO package.\cite{Ganose2018} DOSs and COHPs were generated using LOBSTER and a Gaussian smoothing of 
0.05\,eV was applied to both.\cite{Deringer2011, Maintz2013, Maintz2016} All crystal structures are depicted 
using VESTA.\cite{Momma2011}
	
\section{Acknowledgments}
We are grateful to Professor Si\^an Dutton, Dr. Michelle Johannes, and Professor Fred Wudl for their comments 
and insights. A. K. W. and R. S. acknowledge support from the National Science Foundation through Enabling 
Quantum Leap: Convergent Accelerated Discovery Foundries for Quantum Materials Science, Engineering and 
Information (Q-AMASE-i) award number DMR-1906325. AKC ackowledges support from the Ras Al Khaimah Centre 
for Advanced Materials. The use of computational facilities supported by the National Science Foundation 
(CNS-1725797) and administered by the Center for Scientific Computing (CSC) is gratefully acknowledged. The 
CSC is supported by the California NanoSystems Institute and the Materials Research Science and Engineering 
Center (MRSEC; NSF DMR 2308708) at UC Santa Barbara.

\clearpage

\bibliography{bib-oxides}

\clearpage

\begin{tocentry}
\centering \includegraphics[width=3.0in]{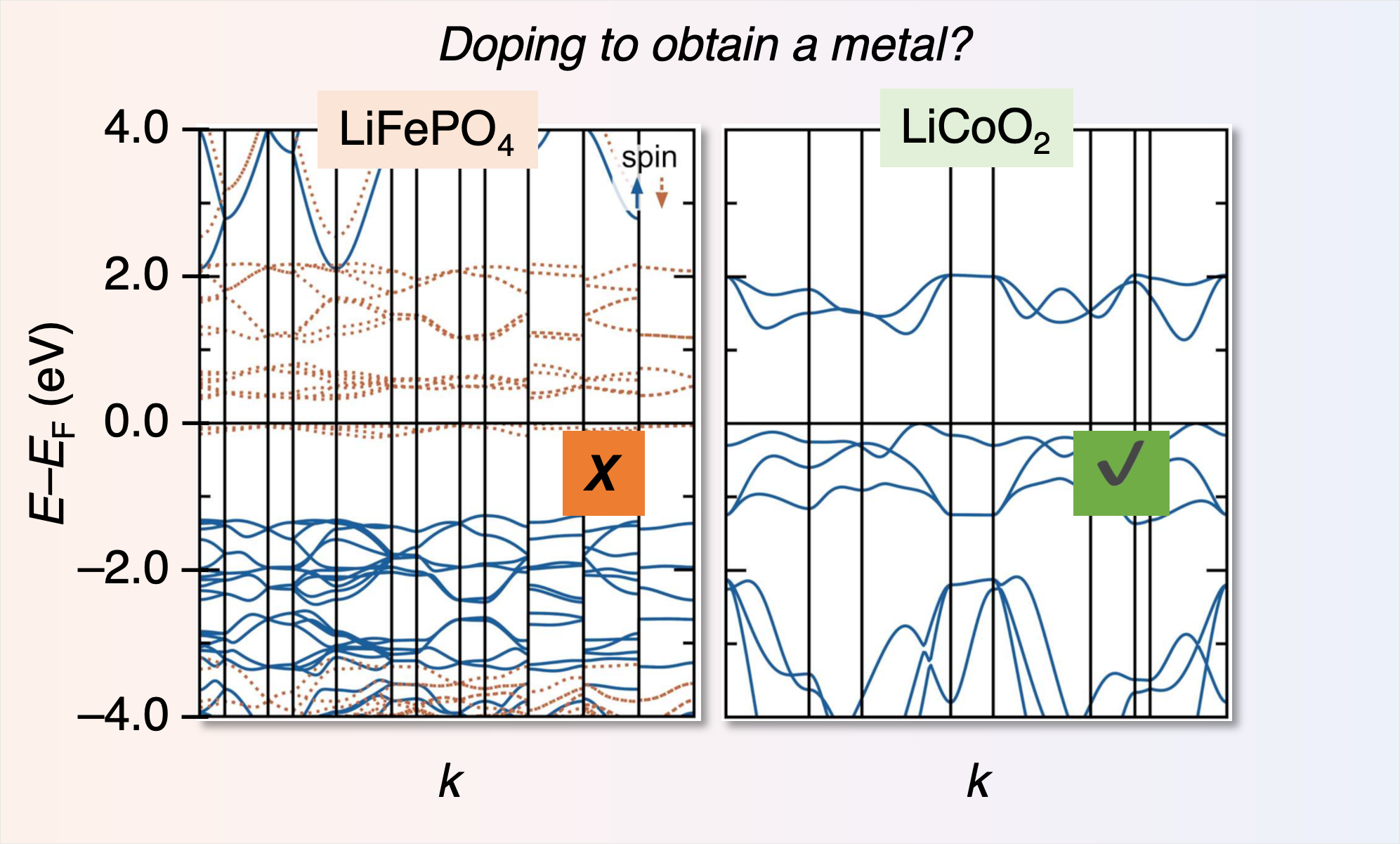}
\end{tocentry}

\end{document}